\documentclass[10pt]{article}

\usepackage{geometry}
\usepackage{array}
\usepackage{longtable}
\usepackage{pdflscape}
\usepackage{url}
\usepackage{amsmath}
\usepackage{amsthm}
\usepackage{bbm}
\usepackage{graphicx}
\usepackage{setspace}
\usepackage{titlesec}
\usepackage{color}
\usepackage{booktabs}
\usepackage[acronym,nonumberlist]{glossaries}
\usepackage{natbib}
\usepackage{caption}
\titleformat{\section}{\bfseries}{\thesection}{2em}{}

\begin{document}

\title{\Large 
	On the Continuous Limit of Weak GARCH} 

\vspace{20pt}

\author{\large Carol Alexander\footnote{University of Sussex Business School, c.alexander@sussex.ac.uk} $\,$ and Emese Lazar\footnote{ ICMA Centre, Henley Business School, Unibversity of Reading, e.lazar@icmacentre.ac.uk}}

\date{ \today}
\maketitle

\onehalfspacing

\thispagestyle{empty}

\begin{abstract}\normalsize
\noindent We prove that the  symmetric weak GARCH limit is a geometric mean-reverting stochastic volatility process with diffusion determined by kurtosis of physical log returns; this provides an improved fit to implied volatility surfaces. When log returns are normal the limit coincides with  Nelson's limit. The limit is unique, unlike strong GARCH limits, because assumptions about  convergence of model parameters is unnecessary -- parameter convergence is uniquely determined by time-aggregation of the weak GARCH process.  \\

\end{abstract}

\vspace{10pt}


\clearpage
\glsresetall
\setcounter{page}{1}

\noindent

\section{Introduction}
\vspace{-10pt}
The symmetric weak GARCH process introduced by Drost and Nijman (1993) is characterised by the absence of parametric conditional distributions of the errors, with the familiar autoregressive equation being defined for the best linear predictor of the residuals rather than the conditional variance. It is the only class of GARCH process which satisfies the time-aggregation property (i.e. doubling or halving the sampling frequency doesn’t change the class, it remains a weak GARCH process). Models that do not satisfy the time-aggregation property may not have unique limit. Nelson (1990) derived a limit of the strong symmetric normal GARCH model as a stochastic variance process with independent Brownians. This is fundamental for limits of other GARCH processes – see Lindner (2009) for a brief overview.  But, like other strong GARCH limits, Nelson’s limit is not unique because it is necessary to make assumptions about the convergence rate of the parameters, and different assumptions lead to different limits which may be either stochastic or deterministic – see Corradi (2000).  Also, because strong GARCH is not time-aggregating, the discretized version of the continuous limit not only depends on the frequency, it may not even be a GARCH process.  

Drost and Werker (1996) introduce continuous-time symmetric GARCH diffusion and jump-diffusion processes that exhibit weak GARCH-type behaviour at all discrete frequencies, showing that their characterisation also depends on the kurtosis of the observed discrete time data. Building on this, here we show that the continuous limit of a symmetric weak GARCH process is a stochastic volatility model similar to Nelson’s limit but the diffusion coefficient is related to the kurtosis of the distribution of log returns. This endows the limit of weak GARCH with an additional parameter and simulations demonstrate how this can improve the fit to implied volatilities when calibrated in the risk-neutral measure. If log returns are normally distributed our limit reduces to Nelson’s strong GARCH diffusion. We also prove that there is no ambiguity about parameter convergence, it follows uniquely and directly from the definition of the weak GARCH process so that the limit is unique. Therefore, within the class of weak GARCH processes, where time aggregation prevails, knowledge of the discrete time GARCH parameters at only one frequency, and knowledge of the kurtosis, completely determines the coefficients of the continuous GARCH process. So, in estimating a continuous time GARCH process in the physical measure it suffices to estimate the discrete time GARCH parameters for the available data frequency. 

\section{The Weak GARCH Process}
\vspace{-10pt}
Following Engle (1982) and Bollerslev (1986) the GARCH(1,1) process for a log return ${y_t} $ can be written as:
 ${y_t} = \mu  + {\varepsilon_t}\quad$with $E\left( {\left. {{\varepsilon _{t + 1}}} \right|{I_t}} \right) = 0, $	
where  ${I_t} $ is the ${\sigma}$-algebra generated by the residual vector  $\left( {{\varepsilon_t}} \right) $. The classical or strong GARCH definition states:
 \begin{equation}\label{eq1}
 E\left( {\left. {\varepsilon _{t + 1}^2} \right|{I_t}} \right) = {h_t},
 \end{equation}
where  ${h_t} $ is the conditional variance. Now in the symmetric version of both strong and weak GARCH,  we assume
${h_t} = \omega  + \alpha \varepsilon_t^2 + \beta {h_{t - 1}} $. But in the weak GARCH process (Drost and Nijman, 1993) $h_t$ is the best linear predictor (BLP) of the squared residuals, not the conditional variance, replacing \eqref{eq1} with:
$$
E\left( {{\varepsilon _{t + 1}}\varepsilon _{t - i}^r} \right) = 0\quad i \ge 0\quad r = 0,1,2; \qquad
E\left( {\left( {\varepsilon _{t + 1}^2 - {h_t}} \right)\varepsilon _{t - i}^r} \right) = 0\quad i \ge 0\quad r = 0,1,2.
$$
The assumption that $0$ and  ${h_t} $ are the BLPs for the residuals and squared residuals respectively, guarantees that the BLP of the squared residuals aggregates in time, but only for symmetric processes. For a finite step-length $\Delta$ we consider the $\Delta$-step process for the residuals and the GARCH process. Time is indexed as $k\Delta$, with $k=1,2, \ldots$ and we use a pre-subscript for the time step and, to be able to compare variances for different step-lengths, we divide by the step-length. Thus ${}_\Delta {h_{k\Delta }} $ denotes the BLP for  ${\Delta ^{ - 1}}{}_\Delta \varepsilon _{k\Delta }^2 $. Using  ${}_\Delta \lambda  = {}_\Delta \alpha  + {}_\Delta \beta  $, for $ i \ge 0$ and $r = 0,1,2$ the annualised weak GARCH process may be written: 
\begin{eqnarray}\label{eq2}
{}_\Delta {y_{k\Delta }} = \Delta \mu  + {}_\Delta {\varepsilon _{k\Delta }},\qquad
{}_\Delta {h_{k\Delta }} = {}_\Delta \omega  + {}_\Delta \alpha {\Delta ^{ - 1}}\varepsilon _{k\Delta }^2 + {}_\Delta \beta {}_\Delta {h_{\left( {k - 1} \right)\Delta }},\\ \nonumber
E\left( {{}_\Delta {\varepsilon _{\left( {k + 1} \right)\Delta }}{}_\Delta \varepsilon _{\left( {k - i} \right)\Delta }^r} \right) = 0, \qquad
E\left( {\left( {{\Delta ^{ - 1}}{}_\Delta \varepsilon _{\left( {k + 1} \right)\Delta }^2 - {}_\Delta {h_{k\Delta }}} \right){}_\Delta \varepsilon _{\left( {k - i} \right)\Delta }^r} \right) = 0.
\end{eqnarray}
The first paper that discusses the continuous limit of GARCH is that of Nelson (1990). Under the conditions:
 $$\omega  = \mathop {\lim }\limits_{\Delta  \downarrow 0} \left( {{\Delta ^{ - 1}}{}_\Delta \omega } \right){\rm{;}}\quad \alpha  = \mathop {\lim }\limits_{\Delta  \downarrow 0} \left( {{\Delta ^{ - 1/2}}{}_\Delta \alpha } \right){\rm{;}}\quad \theta  = \mathop {\lim }\limits_{\Delta  \downarrow 0} \left( {{\Delta ^{ - 1}}\left( {1 - {}_\Delta \lambda } \right)} \right){\rm{;}}\quad 0 < \omega ,\alpha ,\theta  < \infty  $$
the limit will be a stochastic volatility model with independent Brownians, i.e.
$dS_t = \mu \,S_t dt + \sqrt{V_t}\,S_t dB_{1t}$ with
$dV_t = \left( {\omega  - \theta V_t} \right)dt + \sqrt 2 \alpha V_t dB_{2t}$ 
where $V_t$ is the continuous-time limit of  $h_t$. On the other hand, Corradi (2000) proves that, if we assume the following convergence rates:
 $$\omega  = \mathop {\lim }\limits_{\Delta  \downarrow 0} \left( {{\Delta ^{ - 1}}{}_\Delta \omega } \right){\rm{;}}\quad \alpha  = \mathop {\lim }\limits_{\Delta  \downarrow 0} \left( {{\Delta ^{ - 1}}{}_\Delta \alpha } \right){\rm{;}}\quad \theta  = \mathop {\lim }\limits_{\Delta  \downarrow 0} \left( {{\Delta ^{ - 1}}\left( {1 - {}_\Delta \lambda } \right)} \right){\rm{;}}\quad 0 < \omega ,\alpha ,\theta  < \infty  $$
then the continuous-time limit is a deterministic variance model with the same price dynamics but with $
dV_t = \left( {\omega  - \theta V_t} \right)dt
$. 
The difference between the two assumptions lies with the convergence of alpha (at rate  $\sqrt \Delta   $ versus rate $\Delta$). Which assumption is correct has been the subject of considerable debate. Here we argue that the asssumptions of Nelson are correct, but we promote a different continuous limit because it is best to use the time aggregating model. Without time aggregation we have a strong GARCH process for a given frequency, but for any other frequencies the process will not be a strong GARCH process anymore. 

For a weak GARCH using step-lengths $\Delta$ and $\delta$, $\delta < \Delta$, Drost and Nijman (1993) proved the following relationship between the parameters:
 $${}_\Delta \omega  = {}_\delta \omega \left( {1 - {{\left( {{}_\delta \lambda } \right)}^{{\delta ^{ - 1}}\Delta }}} \right){\left( {1 - {}_\delta \lambda } \right)^{ - 1}} \quad \mbox{and} \quad {}_\Delta \alpha  = {\left( {{}_\delta \lambda } \right)^{{\delta ^{ - 1}}\Delta }} - {}_\Delta \beta. $$                                   
 The relationship between the unconditional kurtosis coefficients, denoted  ${}_\Delta \kappa $ and ${}_\delta \kappa  $ respectively, is:
 \begin{equation}\label{eqkurt}
 {}_\Delta \kappa  = 3 + {\Delta ^{ - 1}}\delta \left( {{}_\delta \kappa  - 3} \right) + 6\left( {{}_\delta \kappa  - 1} \right)\frac{{\left( {{\delta ^{ - 1}}\Delta \left( {1 - {}_\delta \lambda } \right) - \left( {1 - {}_\delta {\lambda ^{{\delta ^{ - 1}}\Delta }}} \right)} \right){}_\delta \alpha \left( {1 - {}_\delta {\lambda ^2} + {}_\delta \alpha {}_\delta \lambda } \right)}}{{{{\left( {{\delta ^{ - 1}}\Delta } \right)}^2}{{\left( {1 - {}_\delta \lambda } \right)}^2}\left( {1 - {}_\delta {\lambda ^2} + {}_\delta {\alpha ^2}} \right)}}.
 \end{equation}
Drost and Nijman (1993) derive the following relationship between the low and high frequency parameters:
 $${}_\Delta \beta {\left( {1 + {}_\Delta {\beta ^2}} \right)^{ - 1}} = \left( {{}_{\Delta ,\delta }c {}_\delta \lambda {{\kern 1pt} ^{{\delta ^{ - 1}}\Delta }} - 1} \right){\left( {{}_{\Delta ,\delta }c\left( {1 + {}_\delta \lambda {{\kern 1pt} ^{2{\delta ^{ - 1}}\Delta }}} \right) - 2{\kern 1pt} } \right)^{ - 1}} $$
where 
\begin{eqnarray}\label{eqc}
{}_{\Delta ,\delta }c = \left[ \begin{array}{l}
{\delta ^{ - 1}}\Delta {\left( {1 - {}_\delta \beta } \right)^2} + 2{\delta ^{ - 1}}\Delta \left( {{\delta ^{ - 1}}\Delta  - 1} \right)\left( {1 - {}_\delta \lambda } \right)\left( {1 - {}_\delta \lambda {{\kern 1pt} ^2} + {}_\delta {\alpha ^2}} \right){\left( {{}_\delta \kappa  - 1} \right)^{ - 1}}{\left( {1 + {}_\delta \lambda {\kern 1pt} } \right)^{ - 1}}\\ \nonumber
{\rm{    }} + 4\left( {{\delta ^{ - 1}}\Delta \left( {1 - {}_\delta \lambda {\kern 1pt} } \right) - \left( {1 - {}_\delta \lambda {{\kern 1pt} ^{{\delta ^{ - 1}}\Delta }}} \right)} \right){}_\delta \alpha \left( {1 - {}_\delta \beta {}_\delta \lambda {\kern 1pt} } \right){\left( {1 - {}_\delta \lambda {{\kern 1pt} ^2}} \right)^{ - 1}}
\end{array} \right] \times \\ 
\quad \quad {\left[ {{}_\delta \alpha \left( {1 - {}_\delta \beta {}_\delta \lambda } \right)\left( {1 - {}_\delta \lambda {{\kern 1pt} ^{2{\delta ^{ - 1}}\Delta }}} \right){{\left( {1 - {}_\delta \lambda {{\kern 1pt} ^2}} \right)}^{ - 1}}} \right]^{ - 1}}
\end{eqnarray} 
To derive the continuous limit of this model we are interested in the inverse relationship: expressing the high frequency ($\delta$-step) parameters and their limit based on the low frequency ($\Delta$-step) parameters, for $\delta < \Delta$:
 $${}_\delta \omega  = {}_\Delta \omega \left( {1 - {}_\Delta \lambda {{\kern 1pt} ^{{\Delta ^{ - 1}}\delta }}} \right){\left( {1 - {}_\Delta \lambda } \right)^{ - 1}} \quad \mbox{and} \quad  {}_\delta {\lambda ^{{\delta ^{ - 1}}}} = {}_\Delta \lambda {{\kern 1pt} ^{{\Delta ^{ - 1}}}} $$ Also:
\begin{equation*}
\left( {2\left( {\frac{{{}_\Delta \beta }}{{1 + {}_\Delta {\beta ^2}}}} \right) - 1} \right){}_\delta \alpha \left( {1 - {}_\delta \beta {}_\delta \lambda } \right)\left( {\frac{{1 - {}_\Delta {\lambda ^2}}}{{1 - {}_\delta {\lambda ^2}}}} \right) = \end{equation*}
\begin{equation}\label{eq3}
= \left( {\left( {\frac{{{}_\Delta \beta }}{{1 + {}_\Delta {\beta ^2}}}} \right)\left( {1 + {}_\Delta {\lambda ^2}} \right) - {}_\Delta \lambda } \right)\left( \begin{array}{l}
\Delta {\delta ^{ - 1}}{\left( {1 - {}_\delta \beta } \right)^2} + 2\Delta {\delta ^{ - 1}}\left( {\Delta {\delta ^{ - 1}} - 1} \right){\left( {{}_\delta \kappa  - 1} \right)^{ - 1}}\left( {\frac{{1 - {}_\delta \lambda }}{{1 + {}_\delta \lambda }}} \right)\left( {1 - {}_\delta {\lambda ^2} + {}_\delta {\alpha ^2}} \right)\\
+ 4{\left( {1 - {}_\delta {\lambda ^2}} \right)^{ - 1}}{}_\delta \alpha \left( {\Delta {\delta ^{ - 1}}\left( {1 - {}_\delta \lambda } \right) - \left( {1 - {}_\Delta \lambda } \right)} \right)\left( {1 - {}_\delta \beta {}_\delta \lambda } \right)
\end{array} \right)
\end{equation}

\noindent and
\begin{equation}
\label{eq4}
{}_\delta \kappa  = 1 + \left( {\frac{{\left( {2 + {\delta ^{ - 1}}\Delta \left( {{}_\Delta \kappa  - 3} \right)} \right)\left( {1 - {}_\Delta \lambda {{\kern 1pt} ^{{\Delta ^{ - 1}}\delta }}} \right)\left( {1 - {}_\Delta \lambda {{\kern 1pt} ^{2{\Delta ^{ - 1}}\delta }} + {}_\delta {\alpha ^2}} \right)}}{{\left( {\left( {1 - {}_\Delta \lambda {{\kern 1pt} ^{{\Delta ^{ - 1}}\delta }}} \right)\left( {1 - {}_\Delta \lambda {{\kern 1pt} ^{2{\Delta ^{ - 1}}\delta }} + {}_\delta {\alpha ^2}} \right)} \right) + 6{}_\delta \alpha \left( {1 - {\Delta ^{ - 1}}\delta \left( {1 - {}_\Delta \lambda {\kern 1pt} } \right){{\left( {1 - {}_\Delta \lambda {{\kern 1pt} ^{{\Delta ^{ - 1}}\delta }}} \right)}^{ - 1}}} \right)\left( {1 - {}_\delta \beta {}_\Delta \lambda {{\kern 1pt} ^{{\Delta ^{ - 1}}\delta }}} \right)}}} \right) .\end{equation}	

\section{Continuous Limit of Weak GARCH}
\vspace{-10pt}
The continuous time limit of an econometric model may not offer equivalence with the discrete time model. For equivalence, it must be that the discretization of the continuous limit yields the same discrete time model as the original. Furthermore, the discretized model must be the same for all frequencies. Clearly, this cannot happen if the discrete model does not aggregate in time. Thus, it is only when (1) the original discrete time model is time aggregating, and (2) the model can be discretized at any frequency in the form of the original model, that we have an equivalence between discrete and continuous models. 
The first step for deriving the continuous limit of symmetric weak GARCH is to determine the limits and convergence speeds of the parameters. In contrast to the strong GARCH process, where there is some freedom to choose assumptions about parameter convergence speeds, we now find that it is not needed to make any assumption about parameter convergence. Instead, the time-aggregation property of weak GARCH implies unique convergence speeds for all parameters, as the following proposition shows:\\

\noindent\textbf{Proposition 1:}
The convergence rates for the parameters implied by the weak GARCH model are as follows:
 $$\omega  = \mathop {\lim }\limits_{\Delta  \downarrow 0} {\Delta ^{ - 1}}{}_\Delta \omega ;\quad \alpha  = \mathop {\lim }\limits_{\Delta  \downarrow 0} {\Delta ^{ - 1/2}}{}_\Delta \alpha ;\quad \theta  = \mathop {\lim }\limits_{\Delta  \downarrow 0} {\Delta ^{ - 1}}\left( {1 - {}_\Delta \lambda } \right);\quad 0 < \omega ,\alpha ,\theta  < \infty  $$
Also, the unconditional kurtosis converges to
 $\kappa  = \mathop {\lim }\limits_{\Delta  \downarrow 0} {}_\Delta \kappa  = 3{\left( {1 - {\theta ^{ - 1}}{\alpha ^2}} \right)^{ - 1}} .$ \\
\noindent\textbf{Proof:}
We get  ${}_\delta {\lambda ^{{\delta ^{ - 1}}}} = {}_\Delta {\lambda ^{{\Delta ^{ - 1}}}} $, which is a constant between 0 and 1 denoted  $\exp \left( { - \theta } \right)$ with $\theta > 0.$ Thus  $${}_\Delta \lambda  = \exp \left( { - \theta \Delta } \right) \quad \mbox{and} \quad \mathop {\lim }\limits_{\Delta  \downarrow 0} {\Delta ^{ - 1}}\left( {1 - {}_\Delta \lambda } \right) = \mathop {\lim }\limits_{\Delta  \downarrow 0} {\Delta ^{ - 1}}\left( {1 - \exp \left( { - \theta \Delta } \right)} \right) = \theta  .$$
Also ${}_\delta \omega {\left( {1 - {}_\delta \lambda } \right)^{ - 1}} = {}_\Delta \omega {\left( {1 - {}_\Delta \lambda } \right)^{ - 1}} $ is a positive constant denoted $\omega {\theta ^{ - 1}} $, $\omega >0$ and ${}_\Delta \omega  = \omega {\theta ^{ - 1}}\left( {1 - {}_\Delta \lambda } \right) $, so
$$\mathop {\lim }\limits_{\Delta  \downarrow 0} \left( {{\Delta ^{ - 1}}{}_\Delta \omega } \right) = \omega {\theta ^{ - 1}}\mathop {\lim }\limits_{\Delta  \downarrow 0} {\Delta ^{ - 1}}\left( {1 - \exp \left( { - \theta \Delta } \right)} \right) = \omega  .$$
The formula \eqref{eq4} for kurtosis may now be written:
\begin{equation}\label{eq5}
{}_\delta \kappa  = 1 + \left( {{}_\Delta \kappa  - 3 + 2{\Delta ^{ - 1}}\delta } \right){\left( {{\Delta ^{ - 1}}\delta  + 6\frac{{\left( {{\delta ^{ - 1}}\left( {1 - {}_\Delta \lambda {{\kern 1pt} ^{{\Delta ^{ - 1}}\delta }}} \right) - {\Delta ^{ - 1}}\left( {1 - {}_\Delta \lambda {\kern 1pt} } \right)} \right)}}{{\Delta {{\left( {{\delta ^{ - 1}}\left( {1 - {}_\Delta \lambda {{\kern 1pt} ^{{\Delta ^{ - 1}}\delta }}} \right)} \right)}^2}}}{}_{\Delta ,\delta }A} \right)^{ - 1}} , $$ with
$${}_{\Delta ,\delta }A = \frac{{{}_\delta \alpha {\delta ^{ - 1}}\left( {1 - {}_\Delta \lambda {{\kern 1pt} ^{2{\Delta ^{ - 1}}\delta }}} \right) + {\delta ^{ - 1}}{}_\delta {\alpha ^2} - {\delta ^{ - 1}}{}_\delta {\alpha ^2}\left( {1 - {}_\Delta \lambda {{\kern 1pt} ^{{\Delta ^{ - 1}}\delta }}} \right)}}{{{\delta ^{ - 1}}\left( {1 - {}_\Delta \lambda {{\kern 1pt} ^{2{\Delta ^{ - 1}}\delta }}} \right) + {\delta ^{ - 1}}{}_\delta {\alpha ^2}}} .
\end{equation}
But  $$\mathop {\lim }\limits_{\delta  \downarrow 0} {\delta ^{ - 1}}\left( {1 - {}_\Delta {\lambda ^{{\Delta ^{ - 1}}\delta }}} \right) = \theta \quad \mbox{and} \quad  \mathop {\lim }\limits_{\delta  \downarrow 0} {\delta ^{ - 1}}\left( {1 - {}_\Delta {\lambda ^{2{\Delta ^{ - 1}}\delta }}} \right) = 2\theta  .$$ Thus, using $_\delta \alpha \downarrow 0,$ we have
$$\mathop {\lim }\limits_{\delta  \downarrow 0} {}_{\Delta ,\delta }A = {\left( {2\theta {{\left( {\mathop {\lim }\limits_{\delta  \downarrow 0} \left( {{\delta ^{ - 1}}{}_\delta {\alpha ^2}} \right)} \right)}^{ - 1}} + 1} \right)^{ - 1}} .$$
Hence, taking the limit of \eqref{eq5} as $\delta \downarrow 0$  and then $\Delta \downarrow 0$ yields  $\kappa  = 3{\left( {1 - {\theta ^{ - 1}}\mathop {\lim }\limits_{\delta  \downarrow 0} \left( {{\delta ^{ - 1}}{}_\delta {\alpha ^2}} \right)} \right)^{ - 1}} $.
The limit of the unconditional kurtosis is finite and positive, which forces  $0 \le \mathop {\lim }\limits_{\delta  \downarrow 0} {\delta ^{ - 1}}{}_\delta {\alpha ^2} < \theta  $, so the kurtosis will be higher than 3. To see the speed of convergence for $\alpha$ , we consider the limit $\alpha : = \mathop {\lim }\limits_{\delta  \downarrow 0} {\delta ^{ - w}}{}_\delta \alpha  $ with  $\alpha  \in \left( {0,\infty } \right) $ with $w$ unknown.
Since  $\mathop {\lim }\limits_{\delta  \downarrow 0} {\delta ^{ - 1}}{}_\delta {\alpha ^2} < \theta  $, $w \ge 1/2$, for  $y = \min \left( {w,1} \right)$ and $ z = \min \left( {2w,1} \right) $ we write
\begin{eqnarray*}\label{eq6}
\mathop {\lim }\limits_{\delta  \downarrow 0} {\delta ^{ - y}}\left( {1 - {}_\delta \lambda  + {}_\delta \alpha } \right) = \mathop {\lim }\limits_{\delta  \downarrow 0} {\delta ^{ - y}}\left( {1 - {}_\delta \lambda } \right) + \mathop {\lim }\limits_{\delta  \downarrow 0} {\delta ^{ - y}}{}_\delta \alpha  \in \left( {0,\infty } \right)\\
\mathop {\lim }\limits_{\delta  \downarrow 0} {\delta ^{ - y}}\left( {1 - {}_\delta {\lambda ^2} - {}_\delta \alpha \left( {1 - {}_\delta \lambda } \right) + {}_\delta \alpha } \right) = \mathop {\lim }\limits_{\delta  \downarrow 0} {\delta ^{ - y}}\left( {1 - {}_\delta {\lambda ^2}} \right) + \mathop {\lim }\limits_{\delta  \downarrow 0} {\delta ^{ - y}}{}_\delta \alpha  \in \left( {0,\infty } \right)\\
\mathop {\lim }\limits_{\delta  \downarrow 0} {\delta ^{ - z}}\left( {1 - {}_\delta {\lambda ^2} + {}_\delta {\alpha ^2}} \right) = \mathop {\lim }\limits_{\delta  \downarrow 0} {\delta ^{ - z}}\left( {1 - {}_\delta {\lambda ^2}} \right) + \mathop {\lim }\limits_{\delta  \downarrow 0} {\delta ^{ - z}}{}_\delta {\alpha ^2} \in \left( {0,\infty } \right)
\end{eqnarray*}
Also, using \eqref{eq3}  and noting that  ${}_\delta \kappa  \ne 1 $, since  ${}_\delta {\alpha ^2} > 0 $, we can compute
$$\left( {2\left( {\frac{{{}_\Delta \beta }}{{1 + {}_\Delta {\beta ^2}}}} \right) - 1} \right){}_\delta \alpha \left( {1 - {}_\delta {\lambda ^2} - {}_\delta \alpha \left( {1 - {}_\delta \lambda } \right) + {}_\delta \alpha } \right)\left( {\frac{{1 - {}_\Delta {\lambda ^2}}}{{1 - {}_\delta {\lambda ^2}}}} \right). $$
If  $w > 1/2$, we can multiply the above expression by  ${\delta ^{1 - w - y}} $ and then computing the limit as $\delta$ tends to zero leads to a contradiction in terms of limits. So we must have  $w = 1/2$ and this sets the convergence of $\alpha$. 		\qed \\

\noindent Now consider the conditional variance and the conditional kurtosis of the residuals; where the conditional mean and skewness are equal with zero:
 $${}_\Delta \sigma _{k\Delta }^2 = E\left( {\left. {{\Delta ^{ - 1}}{{\left( {{}_\Delta {\varepsilon _{\left( {k + 1} \right)\Delta }} - \Delta {}_\Delta {\mu _{k\Delta }}} \right)}^2}} \right|{}_\Delta {I_{k\Delta }}} \right) $$ $${}_\Delta {\kappa _{k\Delta }} = E\left( {\left. {{\Delta ^{ - 2}}{}_\Delta \sigma _{k\Delta }^{ - 4}{{\left( {{}_\Delta {\varepsilon _{\left( {k + 1} \right)\Delta }} - \Delta {}_\Delta {\mu _{k\Delta }}} \right)}^4}} \right|{}_\Delta {I_{k\Delta }}} \right) $$
where  ${}_\Delta {I_{k\Delta }} $ is the $\sigma$-algebra generated by the vector  $\left( {{}_\Delta {\varepsilon _{k\Delta }}} \right).$ We divide by $\Delta$ when computing the conditional variance series so that the variance over $\Delta$ is comparable with $\Delta$ times the 1-step variance. 

The conditional expectation of the second moment and the kurtosis must be positive, and we shall assume that the following limits exist for  $k\Delta\le t < (k+1)\Delta$:
 $V\left( t \right): = \mathop {\lim }\limits_{\Delta  \downarrow 0} {_\Delta }{h_t} $    where    $_\Delta {h_t}: = {}_\Delta {h_{k\Delta }} $,   
 $\mu \left( t \right) = \mu $ and
 $\kappa \left( t \right): = \mathop {\lim }\limits_{\Delta  \downarrow 0} {_\Delta }{\kappa_t} $    where    $_\Delta {\kappa_t}: = {}_\Delta {\kappa _{k\Delta }}.$  
 Due to the symmetrical nature of the returns, we can write 
\begin{equation}\label{eq13}
 E\left( {\left. {{\Delta ^{ - 1}}{}_\Delta \varepsilon _{\left( {k + 1} \right)\Delta }^2} \right|{I_{k\Delta }}} \right) = {}_\Delta \sigma _{k\Delta }^2. 
\end{equation}
Note that $ {}_\Delta \sigma _{k\Delta }^2 - {}_\Delta {h_{k\Delta }} $ 
has to be different from zero, otherwise the process will be a semi-strong GARCH, However, we assume that as the time step decreases, the difference between the conditional variance and the BLP of the squared residuals converges to zero at a speed of square root of the time step, i.e. 
 $\mathop {\lim }\limits_{\Delta  \downarrow 0} {\Delta ^{ - 1/2}}\left( {{}_\Delta \sigma_t^2 - {}_\Delta {h_t}} \right) = 0. $ 	
In other words, the BLP of the squared residuals is ‘close’ to the conditional variance process. This is the only assumption we make and we consider that it is non-binding because as the time step decreases, the BLP process becomes more and more informative and so it converges fast to the conditional variance, i.e.
 $V\left( t \right) = \mathop {\lim }\limits_{\Delta  \downarrow 0} {_\Delta }\sigma_t^2 $ where    $_\Delta \sigma_t^2: = {}_\Delta \sigma _{k\Delta }^2 $ for  $k\Delta\le t < (k+1)\Delta, $
so that  $\mathop {\lim }\limits_{\Delta  \downarrow 0} \left( {{}_\Delta \sigma_t^2 - {}_\Delta {h_t}} \right) = 0 $ as well.\\

\noindent\textbf{Theorem 1:}
 The continuous time limit of the weak GARCH process defined in \eqref{eq2} is the following stochastic volatility model, based on the limiting parameters given above and in Proposition 1:
\begin{eqnarray*}
\frac{dS_t}{S_t} &=& \mu dt + \sqrt{V_t}\,dB_{1t},\\
dV_t &=& \left( \omega  - \theta V_t \right)dt + \alpha \sqrt{\left( {\kappa_t - 1} \right)}\,V_t \, dB_{2t},
\end{eqnarray*}	
where  $B_{1t}$ and $B_{2t} $ are independent Brownian motions. \\

\noindent\textbf{Proof:} We employ the convergence theorem for stochastic difference equations to stochastic differential equations given by Nelson (1990). For the returns process we have:
$$E\left( {{\Delta ^{ - 1}}\left. {{y_{\left( {k + 1} \right)\Delta }}} \right|{I_{k\Delta }}} \right) = \mu  + E\left( {{\Delta ^{ - 1}}{\varepsilon _{\left( {k + 1} \right)\Delta }}|{I_{k\Delta }}} \right) = \mu  .$$
And, using \eqref{eq13} it can be shown that:
$E\left( {{\Delta ^{ - 1}}\left. {y_{\left( {k + 1} \right)\Delta }^2} \right|{I_{k\Delta }}} \right) = {h_{k\Delta }} + o\left( 1 \right), $
$$E\left( {{\Delta ^{ - 1}}\left. {\left( {{h_{\left( {k + 1} \right)\Delta }} - {h_{k\Delta }}} \right)} \right|{I_{k\Delta }}} \right) = {\Delta ^{ - 1}}{}_\Delta \omega  - {\Delta ^{ - 1}}\left( {1 - {}_\Delta \lambda } \right){h_{k\Delta }} + \left( {{\Delta ^{ - 1/2}}{}_\Delta \alpha } \right){\Delta ^{ - 1/2}}\left( {\sigma _{k\Delta }^2 - {h_{k\Delta }}} \right) + o\left( 1 \right) $$
and this converges to  $\omega  - \theta V_t $ by Proposition 1. The variance of the variance component is: 
$$E\left( {\left. {{\Delta ^{ - 1}}{{\left( {{h_{\left( {k + 1} \right)h}} - {h_{k\Delta }}} \right)}^2}} \right|{I_{k\Delta }}} \right) = {\Delta ^{ - 1}}{}_\Delta {\alpha ^2}\left( {E\left( {\left. {\left( {{}_\Delta \sigma _{k\Delta }^4} \right){{\left( {{\Delta ^2}{}_\Delta \sigma _{k\Delta }^4} \right)}^{ - 1}}\left( {\varepsilon _{\left( {k + 1} \right)\Delta }^4} \right) - h_{k\Delta }^2} \right|{I_{k\Delta }}} \right)} \right) + o\left( 1 \right). $$
The covariance between the returns and the changes in the variances converges as follows:
$$E\left( {{\Delta ^{ - 1}}\left. {{y_{\left( {k + 1} \right)\Delta }}\left( {{h_{\left( {k + 1} \right)\Delta }} - {h_{k\Delta }}} \right)} \right|{I_{k\Delta }}} \right) = o\left( 1 \right).$$
Therefore, the limits of the expected squared terms and cross-product derived above define the following covariance matrix of the continuous process:
$$A_t = \left( {\begin{array}{*{20}{c}}
	{V_t}&0\\
	0&{{\alpha ^2}\left( {\kappa_t - 1} \right)V{_t^2}}
	\end{array}} \right) .\qed $$

\noindent Discrete-time weak GARCH processes are characterized by (i) the existence of a long-term volatility; (ii) mean reversion in the variance process; (iii) the variance is stochastic, i.e. it has a non-zero variance; and (iv) the variance process is uncorrelated with the returns process, which is an implication of the symmetry of the returns’ distribution, being a requirement of weak GARCH processes. All these properties are also present in the continuous limit above; in addition, in our limit model the variance has a higher variance as compared to the limit of Nelson (1990), which results in extra kurtosis, and can be time-varying. These properties are intuitive and parallel the observed behaviour of implied volatilities in the risk neutral measure: see for example, Bates (1997, 2000) and Bakshi et al. (2003). Note that the limit process reduces to the diffusion derived by Nelson (1990) if  $\kappa  = 3 $, and in this case we obtain the smallest value of the volatility of the variance process, i.e. ${2^{1/2}}\alpha V_t $.  Drost and Werker (1996) postulate that the conditional kurtosis is independent of $t$. However, in our limit this is allowed to be time-varying.  

Finally, we show that there is a discretization of the continuous limit under which the original GARCH model is returned when the series of returns and variances are discretized as follows:\\

\noindent \textbf{Discretization Scheme:}\\
$dt \mapsto  \Delta $, 
$\frac{{dS_t}}{{S_t}}  \mapsto  {}_\Delta {y_{\left( {k + 1} \right)\Delta }}$, 
$V_t\mapsto  {}_\Delta {V_{k\Delta }}$ and 
$dV_t \mapsto {}_\Delta {V_{\left( {k + 1} \right)\Delta }} - {}_\Delta {V_{k\Delta }}
.$
The parameter $\mu$ stays unchanged during discretization. The rest of the parameters are discretized as:
$
\omega  \mapsto  {\Delta ^{ - 1}}{}_\Delta \omega$, 
$\theta  \mapsto {\Delta ^{ - 1}}\left( {1 - {}_\Delta \lambda } \right)$ and 
$\alpha \mapsto {\Delta ^{ - 1/2}}{}_\Delta \alpha 
$
where we specify the parameters  $\left( {{}_\Delta \omega ,{}_\Delta \alpha ,{}_\Delta \beta } \right) $ in terms of the parameters  $\left( {\omega ,\theta ,\alpha } \right) $ of the continuous model:
 ${}_\Delta \omega  = \omega {\theta ^{ - 1}}\left( {1 - \exp \left( { - \theta \Delta } \right)} \right) $ ,	
 ${}_\Delta \lambda  = \exp \left( { - \theta \Delta } \right) $ ,
 ${}_\Delta \alpha  = {}_\Delta \lambda  - {}_\Delta \beta  $ and 
 \begin{equation}\label{eqbeta}
 {}_\Delta \beta  = \frac{1}{2} \frac{{{}_\Delta c\left( {1 + \exp \left( { - 2\theta \Delta } \right)} \right) - 2 + \left( {1 - \exp \left( { - \theta \Delta } \right)} \right){{\left( {{}_\Delta {c^2}{{\left( {1 - \exp \left( { - \theta \Delta } \right)} \right)}^2} - 4{}_\Delta c} \right)}^{1/2}}}}{{{}_\Delta c\exp \left( { - \theta \Delta } \right) - 1}}
 \end{equation}		 
where
\begin{equation}\label{eqc2}
{}_\Delta c = \left[ \Delta \alpha ^2 + 2\alpha ^2\left( \Delta  - \theta ^{- 1}\left( 1 - \exp \left(  - \theta \Delta  \right) \right) \right) + \Delta ^2\theta \left(\theta  - \alpha ^2 \right) \right] \times \left[\frac{1}{2}\alpha ^2 \theta ^{-1} \left( 1 - \exp \left(  - 2\theta \Delta  \right)\right) \right]^{ - 1}.
\end{equation}

\noindent The unconditional kurtosis is discretized as:
 $${}_\Delta \kappa  = 3 + 6\left( {\kappa  - 1} \right){\alpha ^2}\left( {\theta \Delta  - \left( {1 - \exp \left( { - \theta \Delta } \right)} \right)} \right){\theta ^{ - 2}}{\Delta ^{ - 2}}{\left( {{\alpha ^2} + 2\theta } \right)^{ - 1}}$$
 whilst the conditional kurtosis is discretized using $\kappa_t \to {}_\Delta {\kappa _{k\Delta }} = \kappa \left( {k\Delta } \right)\quad {\rm{for}}\quad k\Delta  \le t < \left( {k + 1} \right)\Delta . $ 	
The Brownian motions $B_1$ and $B_2$ that drive the price and variance equations are discretized by assuming a time step of length $\Delta$, and we can express the changes in the Brownian motions at time $t = k\Delta$ as:
\begin{eqnarray*}
{B_1}\left( {\left( {k + 1} \right)\Delta } \right) - {B_1}\left( {k\Delta } \right) = {\Delta ^{1/2}}{}_\Delta {\xi _{\left( {k + 1} \right)\Delta }},\\
{B_2}\left( {\left( {k + 1} \right)\Delta } \right) - {B_2}\left( {k\Delta } \right) = {\Delta ^{1/2}}{}_\Delta {\eta _{\left( {k + 1} \right)\Delta }},
\end{eqnarray*} 
where  ${}_\Delta {\xi _{\left( {k + 1} \right)\Delta }} $is a standard normal variable, ${}_\Delta {\xi _{\left( {k + 1} \right)\Delta }} \, | \, {}_\Delta I_{k\Delta } \sim N\left( {0,1} \right)$ 
and ${}_\Delta {\eta _{\left( {k + 1} \right)\Delta }} $is defined as 
\begin{equation}\label{eqeta}
{}_\Delta {\eta _{\left( {k + 1} \right)\Delta }} = {2^{ - 1/2}}\left( {{}_\Delta \xi _{\left( {k + 1} \right)\Delta }^2 - 1} \right) .
\end{equation}
 Now define the normal variable
 ${}_\Delta {\varepsilon _{\left( {k + 1} \right)\Delta }} = {\Delta ^{1/2}}{}_\Delta V_{k\Delta }^{1/2}{}_\Delta {\xi _{\left( {k + 1} \right)\Delta }}$ and set 
 ${}_\Delta {\tilde \varepsilon _{k\Delta }} = {G^{ - 1}}\left[ {F\left( {{}_\Delta {\varepsilon _{k\Delta }}} \right)} \right],$
where $F$ is the normal distribution and $G$ is the distribution for a variable  ${}_\Delta {\tilde \varepsilon _{k\Delta }} $ that has zero mean and variance  $\Delta {}_\Delta V_{k\Delta }^{} $, like  ${}_\Delta {\varepsilon _{k\Delta }} $, but kurtosis equal to  ${}_\Delta {\kappa _{k\Delta }}.$ This way, the errors of the discretized model  have non-zero excess kurtosis.\\

\noindent \textbf{Discussion:}  The continuous model has two independent sources of randomness yet the discrete model has only one. That is, the discretization reduces the number of sources of randomness in the continuous model, via \eqref{eqeta}. There is no loss of generality using this discretization since the properties of the discretized Brownian motion (mean, variance and correlation) are maintained; ${}_\Delta {\eta _{\left( {k + 1} \right)\Delta }} $ is not exactly normal but it has a zero conditional mean, a unit conditional variance
and zero correlation with  ${}_\Delta {\xi _{\left( {k + 1} \right)\Delta }} $. We are bound to use such a method because, as argued by Lindner (2009, p. 482), the classic discretization does not work in this case.

\noindent\textbf{Theorem 2:}  Under the above discretization scheme the  continuous limit in Theorem 1 returns the original weak GARCH model \eqref{eq2}  and the time aggregation property is preserved. \\

\noindent\textbf{Proof:} The discretization of $\mu dt$ is obvious, and that for $\theta$ and $\omega$  will follow from the discretization of $\omega$ and $\lambda$ because:
$$\omega \Delta  \approx \omega \Delta \frac{{\left( {1 - \exp \left( { - \theta \Delta } \right)} \right)}}{{\theta \Delta }} = {}_\Delta \omega  \quad \mbox{and} \, \, \theta \Delta  \approx 1 - \exp \left( { - \theta \Delta } \right) = 1 - \left( {{}_\Delta \alpha  + {}_\Delta \beta } \right) = 1 - {}_\Delta \lambda.$$
This gives:
${}_\Delta \omega  = \omega {\theta ^{ - 1}}\left( {1 - \exp \left( { - \theta \Delta } \right)} \right) $ and  
${}_\Delta \lambda  = \exp \left( { - \theta \Delta } \right).$ 
As is clear from \eqref{eqbeta} and \eqref{eqc2}, it is the discretization of $\beta$ that is most complex. From the aggregation results in Drost and Nijman (1993) we know that the unconditional kurtosis for a given frequency $\Delta$ may be expressed as a function of the parameters at an arbitrary higher frequency $\delta$ as:
$${}_\Delta \kappa  = 3 + {\Delta ^{ - 1}}\delta \left( {{}_\delta \kappa  - 3} \right) + 6\left( {{}_\delta \kappa  - 1} \right)\frac{{\left( {{\delta ^{ - 1}}\Delta \left( {1 - {}_\delta \lambda } \right) - \left( {1 - {}_\delta {\lambda ^{{\delta ^{ - 1}}\Delta }}} \right)} \right){}_\delta \alpha \left( {1 - {}_\delta {\lambda ^2} + {}_\delta \alpha {}_\delta \lambda } \right)}}{{{{\left( {{\delta ^{ - 1}}\Delta } \right)}^2}{{\left( {1 - {}_\delta \lambda } \right)}^2}\left( {1 - {}_\delta {\lambda ^2} + {}_\delta {\alpha ^2}} \right)}} .$$
Denoting the limit of the unconditional kurtosis by  $\kappa : = \mathop {{\rm{lim}}}\limits_{\delta  \downarrow 0} {}_\delta \kappa  $, we obtain:
\begin{equation}\label{eqkurt2}
{}_\Delta \kappa  = 3 + 6\left( {\kappa  - 1} \right)\frac{{{\alpha ^2}\left( {\theta \Delta  - \left( {1 - \exp \left( { - \theta \Delta } \right)} \right)} \right)}}{{{\theta ^2}{\Delta ^2}\left( {{\alpha ^2} + 2\theta } \right)}}
\end{equation}
where, by Proposition 1, the limit of the unconditional kurtosis is given by  $\kappa  = 3{\left( {1 - {\theta ^{ - 1}}{\alpha ^2}} \right)^{ - 1}} $.  From the proof of Proposition 1, we know that for any two time steps $\Delta> \delta$, $_\Delta\beta$ is the solution to :
\begin{equation}\label{eqbeta2}
\frac{{{}_\Delta \beta }}{{1 + {}_\Delta {\beta ^2}}} = \frac{{{}_{\Delta ,\delta }c{}_\delta {\lambda ^{{\delta ^{ - 1}}\Delta }} - 1}}{{{}_{\Delta ,\delta }c\left( {1 + {}_\delta {\lambda ^{2{\delta ^{ - 1}}\Delta }}} \right) - 2}}
\end{equation}
where  ${}_{\Delta ,\delta }c $ is given by \eqref{eqc}. We want a discretization which ensures that \eqref{eqbeta2} will hold. Taking the limits of \eqref{eqc} when $\delta$ goes to 0, we define:  
$${}_\Delta c: = \mathop {{\rm{lim}}}\limits_{\delta  \downarrow 0} {}_{\Delta ,\delta }c = \left[ {\Delta {\alpha ^2} + {\Delta ^2}\theta \left( {\theta  - {\alpha ^2}} \right) + 2{\alpha ^2}\left( {\Delta  - {\theta ^{ - 1}}\left( {1 - \exp \left( { - \theta \Delta } \right)} \right)} \right)} \right]\left[ \frac{1}{2} \theta ^{ - 1}\alpha ^2\left( {1 - \exp \left( { - 2\theta \Delta } \right)} \right) \right]^{ - 1} .$$
This means that we can discretize the continuous model by solving the following equation:
$$\frac{{{}_\Delta \beta }}{{1 + {}_\Delta {\beta ^2}}} = \frac{{{}_\Delta c\exp \left( { - \theta \Delta } \right) - 1}}{{{}_\Delta c\left( {1 + \exp \left( { - 2\theta \Delta } \right)} \right) - 2}} .$$
First, we have to make sure that this will have solutions, and then we have to show that there is a unique solution between zero and one. Let’s consider the function whose roots we want to find:
$$f\left( x \right) = {x^2} - mx + 1,\quad m = \frac{{{}_\Delta c\left( {1 + \exp \left( { - 2\theta \Delta } \right)} \right) - 2}}{{{}_\Delta c\exp \left( { - \theta \Delta } \right) - 1}} .$$
This has two roots  ${x_1}$ and ${x_2} $ where  ${x_1}{x_2} = 1$ and ${x_1} + {x_2} = m $. If we show that  $m$ is positive, then both roots are positive and one will be less than 1. For the existence we need that $m > 2$. If  ${}_\Delta c\exp \left( { - \theta \Delta } \right) > 1 $ then $m > 2$ is equivalent to  ${\left( {1 - \exp \left( { - \theta \Delta } \right)} \right)^2} > 0 $. Thus, all we need to show is that  ${}_\Delta c\exp \left( { - \theta \Delta } \right) > 1 $, which is equivalent to: 
$6\theta \Delta  + 2{\alpha ^{ - 2}}{\Delta ^2}{\theta ^3} + 5\exp \left( { - \theta \Delta } \right) > \exp \left( {\theta \Delta } \right) + 2{\Delta ^2}{\theta ^2} + 4 .$ Both sides of the above equation converge to 5 when $\Delta \rightarrow 0$, and it can be shown, using derivatives with respect to  $\Delta  $, that the left hand side converges faster. Thus  ${}_\Delta c\exp \left( { - \theta \Delta } \right) > 1 $, so for any small step $\Delta$ close enough to zero there will always be a unique solution for $_\Delta \beta$  between zero and one that satisfies the above equation; this solution will be: 
$${}_\Delta \beta  = \frac{1}{2} \left( {m - {{\left( {{m^2} - 4} \right)}^{1/2}}} \right) = \frac{{{}_\Delta c\left( {1 + \exp \left( { - 2\theta \Delta } \right)} \right) - 2 + \left( {1 - \exp \left( { - \theta \Delta } \right)} \right){{\left( {{}_\Delta {c^2}{{\left( {1 + \exp \left( { - \theta \Delta } \right)} \right)}^2} - 4{}_\Delta c} \right)}^{1/2}}}}{{2\left( {{}_\Delta c\exp \left( { - \theta \Delta } \right) - 1} \right)}}.$$
Also, we have that ${}_\Delta \alpha  = \exp \left( { - \theta \Delta } \right) - {}_\Delta \beta  $ . The discretization of the Brownian motions in our scheme is obvious, whilst there is no loss of generality in assuming \eqref{eqeta}.
Now, we have  ${y_{\left( {k + 1} \right)\Delta }} = \mu \Delta  + {}_\Delta {\varepsilon _{\left( {k + 1} \right)\Delta }} $, hence  $E\left( {{\Delta ^{ - 1}}{}_\Delta \varepsilon _{\left( {k + 1} \right)\Delta }^2| {{}_\Delta {I_{k\Delta }}} } \right) = {}_\Delta {V_{k\Delta }} $ with:
$${}_\Delta {V_{\left( {k + 1} \right)\Delta }} = {}_\Delta \omega  + {}_\Delta \alpha {\Delta ^{ - 1}}{}_\Delta \tilde \varepsilon _{\left( {k + 1} \right)\Delta }^2 + {}_\Delta \beta {}_\Delta {V_{k\Delta }} + {}_\Delta {u_{\left( {k + 1} \right)\Delta }}, $$ 
$$\\{}_\Delta {u_{\left( {k + 1} \right)\Delta }}= {}_\Delta \alpha {}_\Delta {V_{k\Delta }}\left[ {1 - {{\left( {{2^{ - 1}}\left( {{}_\Delta {\kappa _{k\Delta }} - 1} \right)} \right)}^{1/2}} + \left( {{{\left( {{2^{ - 1}}\left( {{}_\Delta {\kappa _{k\Delta }} - 1} \right)} \right)}^{1/2}} - 1} \right){}_\Delta \tilde \xi _{\left( {k + 1} \right)\Delta }^2} \right] $$
where  ${}_\Delta \tilde \xi _{\left( {k + 1} \right)\Delta }^{}$ has an unconditional kurtosis of  ${}_\Delta {\kappa _{\left( {k + 1} \right)\Delta }} $, which can be approximated by  ${}_\Delta {\kappa _{k\Delta }} $.
So far we have considered the conditional variance; for the BLP of the squared residuals we have:
$${}_\Delta {h_{k\Delta }} = {}_\Delta {V_{k\Delta }} - \sum\limits_{j = 0}^k {{}_\Delta {\beta ^j}{}_\Delta {u_{\left( {k - j} \right)\Delta }}}  .$$
It is easy to see that this follows a GARCH process as ${}_\Delta {h_{k\Delta }} = {}_\Delta \omega  + {}_\Delta \alpha \left( {{\Delta ^{ - 1}}{}_\Delta \varepsilon _{k\Delta }^2} \right) + {}_\Delta \beta {}_\Delta {h_{\left( {k - 1} \right)\Delta }} $.
To have a weak GARCH we have to show that  ${}_\Delta {h_{k\Delta }} $ is the BLP of  ${\Delta ^{ - 1}}{}_\Delta \tilde \varepsilon _{\left( {k + 1} \right)\Delta }^2 $ , which requires showing that:
$E\left( {\left( {{\Delta ^{ - 1}}{}_\Delta \tilde \varepsilon _{\left( {k + 1} \right)\Delta }^2 - {}_\Delta {h_{k\Delta }}} \right){}_\Delta \tilde \varepsilon _{\left( {k - i} \right)\Delta }^r} \right) = 0$ for $i \ge 0$, $r = 0,1,2 $. Since $E\left( {\left( {{}_\Delta {V_{\left( {k - j - 1} \right)\Delta }} - {\Delta ^{ - 1}}{}_\Delta \tilde \varepsilon _{\left( {k - j} \right)\Delta }^2} \right){}_\Delta \tilde \varepsilon _{\left( {k - i} \right)\Delta }^r} \right) = 0$, this reduces to showing $ E\left( {{}_\Delta {u_{\left( {k - j} \right)\Delta }}{}_\Delta \tilde \varepsilon _{\left( {k - i} \right)\Delta }^r} \right) = 0$.
This is satisfied for $i \ne j$. We now show the proof for  $r = 1$ and $i = j$: We have to show that:
$$E\left( {\left( {{}_\Delta {V_{\left( {k - i - 1} \right)\Delta }} - {\Delta ^{ - 1}}{}_\Delta \tilde \varepsilon _{\left( {k - i} \right)\Delta }^2} \right){}_\Delta \tilde \varepsilon _{\left( {k - i} \right)\Delta }^{}} \right) = 0,\quad i \ge 0 $$ or
$$E\left( {E\left( {{}_\Delta {V_{\left( {k - i - 1} \right)\Delta }}{}_\Delta \tilde \varepsilon _{\left( {k - i} \right)\Delta }^{} - {\Delta ^{ - 1}}{}_\Delta \tilde \varepsilon _{\left( {k - i} \right)\Delta }^3} \right)| {{}_\Delta {I_{\left( {k - i - 1} \right)\Delta }}} } \right) = 0,\quad i \ge 0, $$
which is clearly true. Also:
$E\left( {\left. {{}_\Delta \tilde \varepsilon _{\left( {k - i} \right)\Delta }^2} \right|{}_\Delta {I_{\left( {k - i - 1} \right)\Delta }}} \right) = \Delta {}_\Delta {V_{\left( {k - i - 1} \right)\Delta }} $,   	
and $E\left( {\left. {{}_\Delta \tilde \varepsilon _{\left( {k - i} \right)\Delta }^4} \right|{}_\Delta {I_{\left( {k - i - 1} \right)\Delta }}} \right) = {\Delta ^2}{}_\Delta V_{\left( {k - i - 1} \right)\Delta }^2{}_\Delta {\kappa _{\left( {k - i - 1} \right)\Delta }} $. Thus, we have a weak GARCH specification; this means that the time aggregation is preserved by our discretization. It is easy to see that ${}_\Delta {\lambda ^{{\Delta ^{ - 1}}}} = {}_\delta {\lambda ^{{\delta ^{ - 1}}}} $and that  ${}_\Delta \omega  = {}_\delta \omega \left( {1 - {}_\delta {\lambda ^{{\delta ^{ - 1}}\Delta }}} \right)\left( {1 - {}_\delta \lambda } \right) $. We also have the relations \eqref{eqkurt2} for the kurtosis and \eqref{eqbeta2} for $\beta$. For the kurtosis, we need to prove \eqref{eqkurt}, that is:
$$\left( {\kappa  - 1} \right){\alpha ^2}{\left( {1 - {}_\delta \lambda } \right)^2}\left( {1 - {}_\delta {\lambda ^2} + {}_\delta {\alpha ^2}} \right) = \left( {2{\delta ^2}{\theta ^2}\left( {{\alpha ^2} + 2\theta } \right) + 6\left( {\kappa  - 1} \right){\alpha ^2}\left( {\theta \delta  - \left( {1 - {}_\delta \lambda } \right)} \right)} \right){}_\delta \alpha \left( {1 - {}_\delta \lambda  + {}_\delta \alpha {}_\delta \lambda } \right) .$$
After some algebra, this may be written as: 
$$\frac{{{}_\delta \beta }}{{1 + {}_\delta {\beta ^2}}} = \frac{{\left( {\delta {\alpha ^2} + {\delta ^2}\theta \left( {\theta  - {\alpha ^2}} \right) + 2{\alpha ^2}\left( {\delta  - \left( {1 - {}_\delta \lambda } \right)/\theta } \right)} \right){}_\delta \lambda  - \raise.5ex\hbox{ $\scriptstyle 1 $}\kern-.1em/
		\kern-.15em\lower.25ex\hbox{ $\scriptstyle 2 $} {\alpha ^2}{\theta ^{ - 1}}\left( {1 - {}_\delta {\lambda ^2}} \right)}}{{\left( {\delta {\alpha ^2} + {\delta ^2}\theta \left( {\theta  - {\alpha ^2}} \right) + 2{\alpha ^2}\left( {\delta  - \left( {1 - {}_\delta \lambda } \right)/\theta } \right)} \right)\left( {1 + {}_\delta {\lambda ^2}} \right) - {\alpha ^2}{\theta ^{ - 1}}\left( {1 - {}_\delta {\lambda ^2}} \right)}}.$$
Since the above expression holds, we have shown that the kurtosis is time aggregating.

\section{Simulation of the Model}
\vspace{-10pt} 
We simulate the continuous GARCH process 100,000 times, with 1000 steps, computing the price of a standard European option, and then finding the Black-Scholes (1973) implied volatility. Figure 1 (a) compares the volatility skew based on Nelson$'$s diffusion (solid line), with those from the weak GARCH diffusion, showing how different values of instantaneous kurtosis (assumed constant) can influence the shape of the implied volatility. When the instantaneous kurtosis is time dependent, in our case  $\kappa \left( {T - t} \right) = {\rm{7}} - {\rm{2(}}T - t{\rm{)}} $, the steepness of the weak GARCH skew decreases with the option maturity, as in Figure 1 (b), showing that the GARCH limit has considerable flexibility to fit a volatility smile surface through a suitable parameterization of the instantaneous kurtosis.\\

\noindent\textbf{Figure 1:} Comparison of volatility smiles generated by of the weak and strong GARCH diffusions: $\mbox{(a)} \, T-t =1; \, \mbox{(b)} \,  T \, \mbox{varies}$ with $
\omega  = 0.0045;\, \alpha  = 0.1; \, \theta  = 0.05;\, \mu  = \tau  = r = 0;\, \sigma_0 = 30\%$
 \begin{figure}[h!]
	\begin{centering}
		\includegraphics[width=0.9\textwidth]{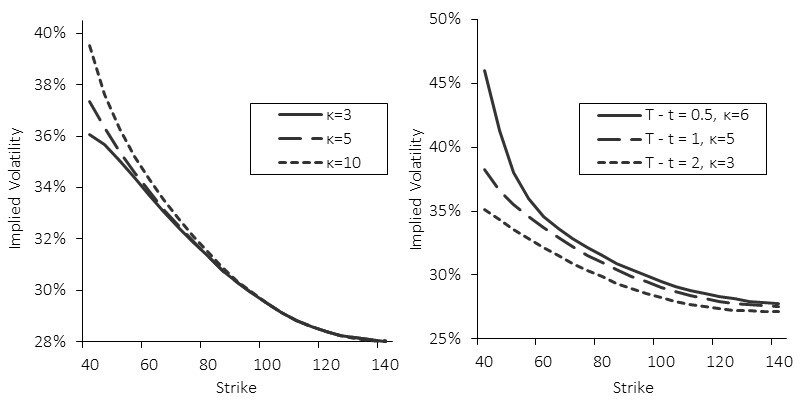}
		\end{centering}
\end{figure}
\vspace{-20pt}
\section{Conclusions}
\vspace{-10pt}
We have presented several arguments which motivate the use of the weak rather than the strong version of the model for deriving a weak limit, i.e. a limit in distribution. There are four problems with the strong GARCH: First, it is not time aggregating: if we generate a GARCH process and then resample at another frequency the result is not a GARCH process. Second, the limit of strong GARCH may only be derived by making a specific assumption about the convergence of the parameters and different assumptions lead to different limits; Third, any discretization of the strong GARCH diffusion is not a GARCH model. And fourth, the variance of the variance is either zero or too small to fit the implied skew. This paper has derived the continuous limit of the weak GARCH by conjecturing only that the difference between the GARCH BLP process and the conditional variance converges to zero with the square root of the step-length. This GARCH model is time aggregating and it implies the convergence rates for all parameters (no need to make assumptions about these). Furthermore, the limit model derived is unique and a discretization that returns the original weak GARCH model is given. The weak GARCH diffusion is a stochastic variance process with independent Brownian motions in which the variance diffusion coefficient is related to the instantaneous kurtosis,  and the limit reduces to Nelson’s GARCH diffusion when the excess kurtosis is zero. \\

\newpage \singlespacing 
\noindent\textbf{References}\\ \\
Badescu A., Elliott R.J., Ortega J-P (2014) `Quadratic Hedging Schemes for Non-Gaussian GARCH Models', Journal of Economic Dynamics and Control Vol. 42, 13-32.\\ 
Badescu A., Elliott R.J., Ortega J-P (2015) `Non-Gaussian GARCH Option Pricing Models and their Diffusion Limits', European Journal of Operational Research Vol. 247, 820-830.\\ 
Badescu A., Cui. Z., Ortega J-P (2017) `Non-Gaussian GARCH Option Pricing Models, Variance-Dependent Kernels and Diffusion Limits', Journal of Financial Econometrics, Vol. 15, 602–648.\\
Bollerslev, T. (1986) `Generalized Autoregressive Conditional Heteroskedasticity', Journal of Econometrics Vol. 31, 309-328.\\
Brown, L. D., Y. Wang and L.H. Zhao (2002) `On the Statistical Equivalence at Suitable Frequencies of GARCH and Stochastic Volatility Models with Suitable Diffusion Models', University of Pennsylvania Working paper.\\
Buchmann, B. and Müller G. (2012) `Limit Experiments of GARCH', Bernoulli Vol. 18, 64-99.\\
Corradi, V. (2000) `Reconsidering the Continuous Time Limit of the GARCH(1,1) Process', Journal of Econometrics Vol. 96, 145-153.\\
Drost, F.C. and Nijman, T.E. (1993) `Temporal Aggregation of GARCH Processes', Econometrica Vol. 61 (4), 909-927.\\
Drost, F.C. and Werker, B.J.M. (1996) `Closing the GARCH Gap Continuous Time GARCH Modelling', Journal of Econometrics Vol. 74, 31-57.\\
Engle, R.F. (1982) `Autoregressive Conditional Heteroscedasticity with Estimates of the Variance of United Kingdom Inflation', Econometrica Vol. 50 (4), 987-1007.\\
Fornari, F. and A. Mele (2005) `Approximating Volatility Diffusions with CEV-ARCH Models', Journal of Economic Dynamics and Control Vol. 30 (6), 931-966.\\
Kallsen J. and Vesenmayer B. (2009) `CO-GARCH as a Continuous-Time Limit of GARCH(1,1)', Stochastic Processes and their Applications Vol. 119, 74-98.\\ 
Kluppelberg, C., Lindner, A. and Maller, R. (2004) `A Continuous Time GARCH Process Driven by Levy Process Stationarity and Second Order Behaviour', Journal of Applied Probability Vol. 43 (3), 601-622.\\
Lindner A. (2009) `Continuous Time Approximations to GARCH and Stochastic Volatility Models', Handbook of Financial Time Series, Edited by Andersen T., Davis R., Kreiss, J-P. and Mikosch, T. (Springer), 481-496.\\
Maller, R., Miller, G. and Szimayer, A. (2008) `GARCH Modelling in Continuous Time for Irregularly Spaced Time Series Data'. Bernoulli, Vol. 14, 519-542.\\
Meddahi, N. and Renault, E. (2004) `Temporal Aggregation of Volatility Models', Journal of Econometrics 19, 355 - 379.\\
Mele, A. and F. Fornari (2000) `Stochastic Volatility in Financial Markets Crossing the Bridge to Continuous Time', Kluwer Academic Publishers.\\
Muller, U.A., M.M. Dacorogna, R. Davé, R.B. Olsen, O.V. Pictett and J.E. Von Weizsäcker (1997) `Volatilities of Different Time Resolutions – Analyzing the Dynamics of Market Components', Journal of Empirical Finance Vol. 4, 213-239.\\
Nelson, D.B. (1990) `ARCH Models as Diffusion Approximations', Journal of Econometrics Vol. 45, 7-38.\\
Trifi, A. (2006) `Issues of Aggregation Over Time of Conditional Heteroscedastic Volatility Models What Kind of Diffusion Do We Recover?', Studies in Nonlinear Dynamics and Econometrics Vol. 10 (4), 1314-1323.\\
Wang, Y. (2002) `Asymptotic Non-Equivalence of GARCH Models and Diffusions', The Annals of Statistics Vol. 30, 754-783.\\
Zhang, R.-M. and Lin, Z.-Y. (2012) `Limit theory for a geneeral class of GARCH models with just barely infinite variance', Journal of Time Series Analysis Vol. 33, 161-174\\
Zheng, Z. (2005) `Re-crossing the Bridge from Discrete Time to Continuous Time Towards a Complete Model with Stochastic Volatility I', available at SSRN http//ssrn.com/abstract=694261.

\end{document}